\documentclass{article}

\usepackage{arxiv}

\usepackage[utf8]{inputenc} 
\usepackage[T1]{fontenc}    
\usepackage{hyperref}       
\usepackage{url}            
\usepackage{booktabs}       
\usepackage{amsfonts}       
\usepackage{nicefrac}       
\usepackage{microtype}      
\usepackage{lipsum}
\usepackage[authoryear]{natbib}
\usepackage{ifpdf}
\usepackage[T1]{fontenc}
\usepackage{sourcesanspro}
\usepackage{newtxmath}
\usepackage{textcomp}%
\usepackage{xcolor}%
\usepackage{hyperref}
\usepackage{comment}
\usepackage{graphicx}
\title{Targeted particle delivery via vortex ring reconnection}

\author{
 Joseph Mouallem$^{1,2}$,  Joshua Wawryk$^{1}$, Hamid Daryan$^{1}$, Zhao Pan$^{1}$, Jean-Pierre Hickey$^{1}$\\
  $^{1}$Department of Mechanical and Mechatronics Engineering, University of Waterloo \\ $^{2}$Atmospheric and Oceanic Sciences,Princeton University
}

\begin{document}
\maketitle

\begin{abstract}
A conceptual model for targeted particle delivery is proposed using controlled vortex ring reconnection. Entrained particles can be efficiently transported within the core of the vortex ring which is propelled via self-induction. A pair of these particle-transporting vortices travelling in the streamwise direction along parallel trajectories will mutually interact resulting in vortex reconnection. The reconnection causes a topological change to the vortex rings which is accompanied by a rapid repulsion in a perpendicular plane to the direction of travel; effectively, transporting the particles to the desired delivery site on the sidewalls. This work explains the physics of the process and the considerations for targeted delivery.
\end{abstract}


\section{Introduction}
The targeted delivery of dispersed particles is encountered in a number of applications such as cold spray, material coatings, micro-organism deposition, and targeted drug delivery. In many of these applications, the solid or liquid particle transport and delivery is done through the control of a carrier fluid to a deposition site. Fluid-based particle transport is longstanding concept as seen in the 'tea leaf paradox' elucidated by \citet{einstein1926}. Modern fluid-controlled particle delivery can be accomplished by generating secondary flows through a modification of the pressure field  \citep{marzo2018acoustic,Guo2020}, using swirl vortex breakdown to direct particles \citep{Kang2020}, or via particle-laden ring vortices for ophthalmic medication delivery \citep{herpin2017precision}. These emerging particle-delivery concepts offer limited directional control in more generic internal flow setups, for example, where delivery to a specific location along the sidewalls of an internal channel flow is desired. The need for targeted particle deposition within internal flows arises in applications such as internal surface treatments or drug delivery in the human airways. 

To address the shortcomings in internal flow particle delivery, we propose a versatile  targeted particle deposition approach using a  pair of reconnecting, particle-laden, laminar vortex rings.  A pair of side-by-side vortex rings, with particles loaded in the vortex core, advance through self-induction via Biot-Savart; the vortex behaves as a compact solitary wave with translational motion, thus transporting the particles.  If particles are located and remain within the vortex core, they can travel a much larger distance compared to transport in a uniform flow \citep{hunt2007vortices}. As the co-aligned rings approach and/or diffuse into each other,  mutual induction will cause the rings to reconnect, thus transferring circulation--and particle momentum--to the plane perpendicular to the initial direction of travel. The reconnection provides an effective means for particle delivery to the internal sidewalls of a channel. 

Viscous vortex reconnection represents a mathematical singularity in the incompressible Navier-Stokes equations and is the only known event resulting in a topological change in fluid dynamics \citep{moffatt2019}.  As two anti-parallel vortices approach each other, either by mutual induction or an imposed collision velocity, circulation is annihilated via viscous cross-diffusion and transferred to the perpendicular plane. The circulation transfer via reconnecting vortex lines results in such a rapid and violent repulsion that it is a  significant source of hydrodynamic noise \citep{Daryan2020}.  Numerous configurations of vortex ring reconnection have been studied, although of interest is the configuration of two side-by-side rings of identical size and circulation, on initially parallel trajectories \citep{Kida1991}. In this configuration, the pair of rings begin by translating on parallel trajectories due to their self-induced motion, eventually reconnecting via mutual-induction.
Particle-laden vortical flows have been used for directional mass delivery, dispersion, and mixing \citep[e.g.][]{Yagami2011,herpin2017precision,zhang_rival_2020}. In canonical vortical flows, the particle transport  within the vortex can be approximated by the local induction equation. An important factor for effective particle transport lies in the ability of the inertial particles to remain within the vortex core during the translational motion. The radius of the vortex core as well as the particle Stokes number affect the residence time of a particle within the vicinity of a vortex \citep{Fung2000}. 

Here, we propose an approach for targeted particle delivery towards the sidewalls of an internal flow. The approach combines the effective translational mass transport of inertial particles within vortex rings with a series of tuned vortex reconnection events for targeted sidewall delivery. The vortex circulation, ring size, relative angle, vortex separation, and bulk fluid velocity represent independent variables that can be tuned for targeted particle deposition. This work elucidates the general approach and investigates key parameters for the particle delivery via Eulerian-Lagrangian direct numerical simulations (DNS). The approach is suited for aerosolized drug delivery or targeted internal flow surface coatings.

\section{Methodology}
\subsection{Numerical details}
Point-particle DNS of a side-by-side pair of vortex rings within a channel are performed using Pencil, a high-order finite-difference code on a fully-structured, Cartesian mesh \citep{Pencil2011}. A sixth-order central, compact finite difference scheme is used for spatial discretization while the temporal integration is performed via a third-order Runge-Kutta scheme. An Eulerian-Lagrangian modeling approach is adopted: the fluid phase is considered as a continuum, while the solid phase is described as discrete point-particles within the fluid phase. The Navier-Stokes equations, in fully compressible form, are used to describe the Eulerian flow field (denoted with subscript $g$) whereas the Lagrangian particles  (denoted with subscript $p$) obey classical Newtonian physics where only the local particle drag, computed via Schiller-Naumann correlation \citep{schiller1933}, is considered. 

\subsection{Simulation setup}
Numerical simulations were performed in a temporally-evolving channel with a square cross-section. The computational domain size is ($4\pi$, $2\pi$, $2\pi$) in $x$, $y$ and $z$, respectively. We consider $x$ to be the streamwise direction with periodic boundary conditions; no-slip wall conditions are imposed in the other directions. The domain was discretized using $512\times256\times256$ uniformly distributed grid points. This was deemed sufficient to capture all scales of the flow evolution at this low Reynolds number, including the vortex reconnection process. The flow is initialized with two identical vortex rings with an outer radius $R_o=\frac{\pi}{4}$ (defined here as the distance from the center of the vortex ring to the center of the vortex core); the vortex core radius was set to $\sigma=0.2R_o$. The vortex rings are aligned side-by-side with their centerline axes parallel and aligned to the $x$-direction.  The centers of the circular rings are offset respectively by $-(R_o + \sigma)$ and $R_o + \sigma$ in the $z$ direction off from the center of the $yz$ plane. By increasing the initial separation between the vortex rings, we can delay the start of the first reconnection event, thus allowing farther streamwise transport of the particles within the channel. In the present work we consider two cases with an initial separation, or gap, equivalent to one and two vortex core radius, $\sigma$, in the $z$ direction.  The magnitude of the initial circulation of each vortex is set to $\Gamma=1$ and the vorticity distribution follows:  
$ \omega_\theta = \frac{\Gamma}{2\pi \sigma^2} \exp\left({\frac{-r^2}{2\sigma^2}}\right)$, where  $\omega_\theta$ is the azimuthal vorticity and $r$ is the distance to the vortex core center. From the given vorticity distribution, a Poisson's equation is solved to set the initial velocity distribution. Additionally, cases with non-zero base flow velocity are considered; in these cases the vortex rings are superposed over a fully developed, laminar, low-Reynolds number velocity profile for rectangular ducts \citep{Shah1978}. For cases with a background flow, we set the value of the centerline flow to be a percentage of the maximum velocity magnitude in the vortex rings. 
Simulations are performed for different particle Stokes numbers. The particle Stokes number $St$ is defined as the ratio of the particle momentum response time to a characteristic flow time scale $St=\tau_p/\tau_g$. The particle response time is given by ${\tau_{p}={d^2_p\rho_p}/{18\mu_g}}$, where $d_p$, $\rho_p$, and $\mu_g$ are respectively the particle diameter, density, and gas viscosity. The flow time scale represents the ratio of the vortex outer diameter to the vortex circulation ($\tau_g=L^2/\Gamma$).

\begin{figure}
    \centering
    \includegraphics[width=0.7\textwidth]{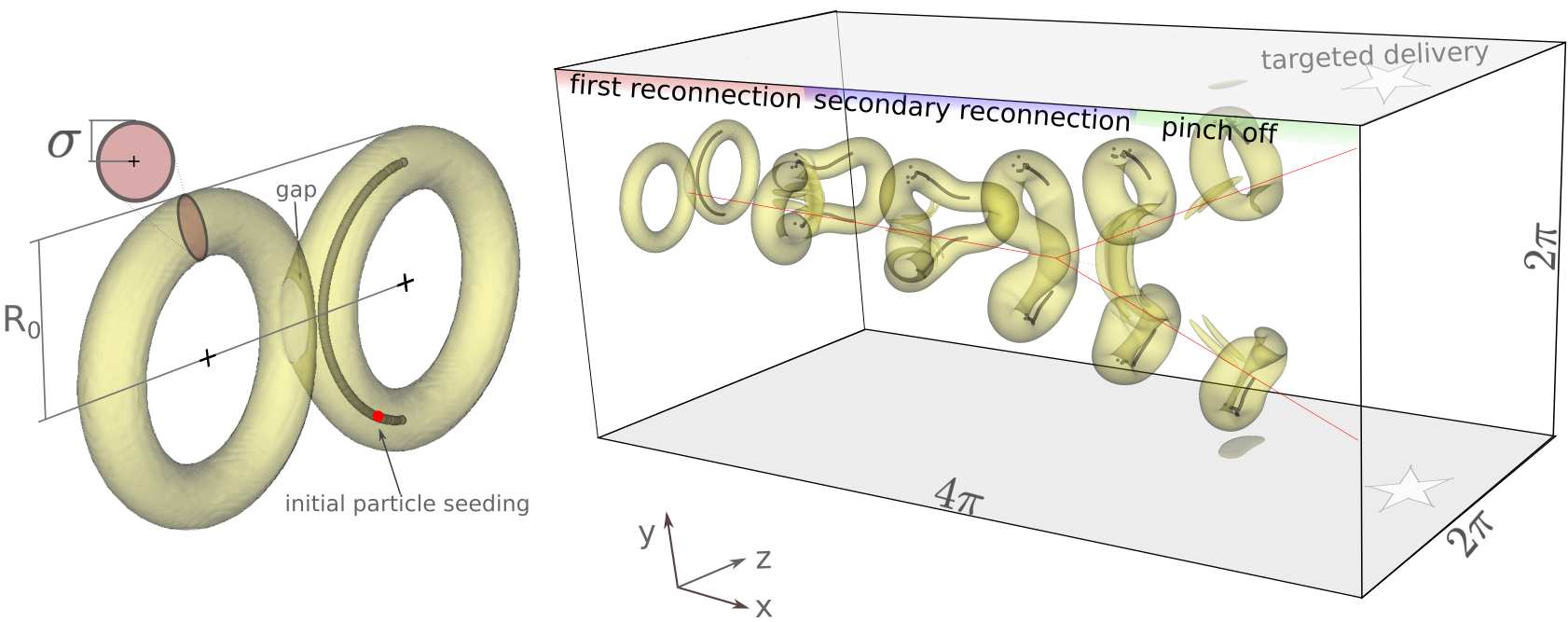}
    \caption{Initial particle-laden vortex ring configuration (left) and evolution of the successive reconnection events, streamwise translated for clarity (right). The isosurfaces of enstrophy  (transparent yellow) are overlaid on the particles (small, grey spheres). Not to scale}
    \label{fig:vortex_reconnection_with_particles}
\end{figure}

\section{Results and Discussion}
The qualitative reconnection process of side-by-side vortex rings is illustrated in figure~\ref{fig:vortex_reconnection_with_particles}. As the vortex rings approach each other, their mutual induction enhances the collision and causes a topological change in the individual vortex lines. This is followed by a strong repulsion of the bridges leading to the change in polarity and large curvature of newly reconnected vortex lines. The strong repulsion of the vortex lines with the conservation of angular momentum causes the outer edges of the rings to bow inwards leading to a second reconnection event. This second reconnection generates two rings that pinch off and travel towards the top and bottom walls. The vortex reconnection is a well-established phenomena, interested readers can consult \citep{wu2007vorticity} for further details.

\begin{figure}
    \centering
    \includegraphics[width=0.8\textwidth]{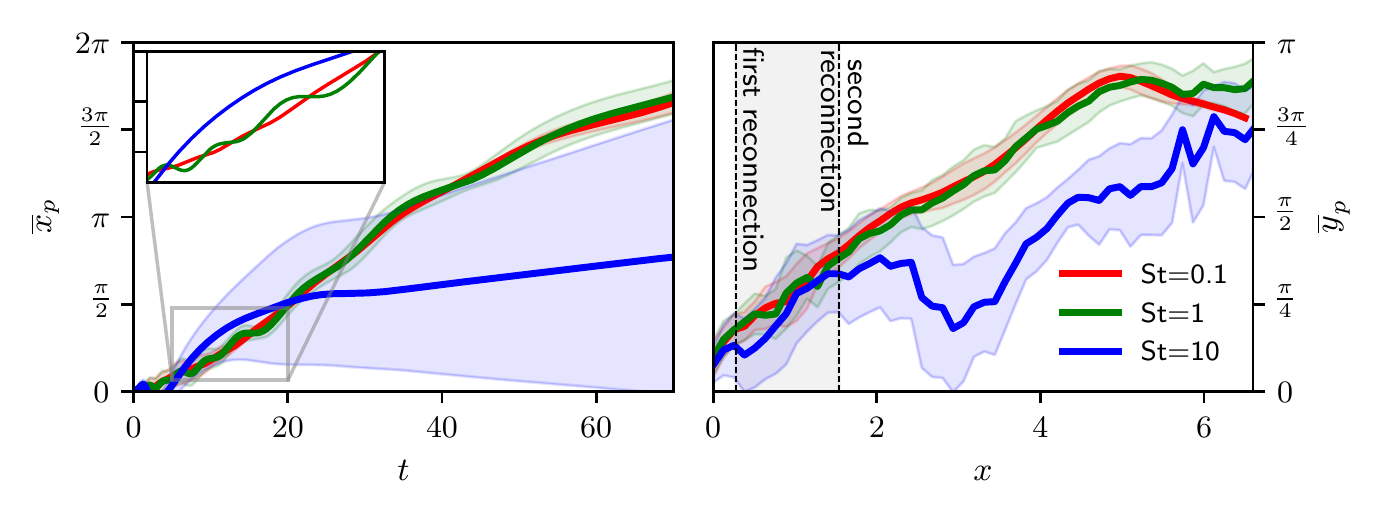} \\
    \vspace{-0.32cm}
    \includegraphics[width=0.365\textwidth]{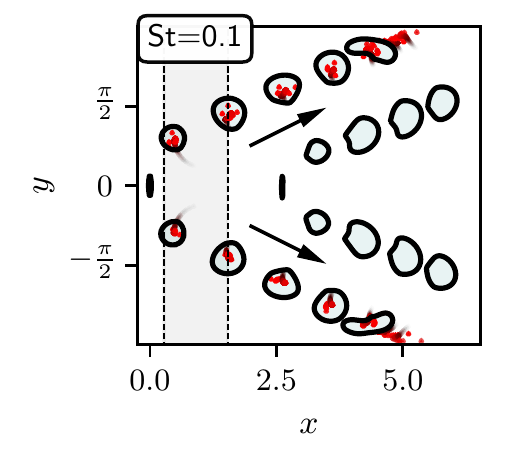}
    \includegraphics[width=0.28\textwidth,trim=32.25 0 0 0, clip]{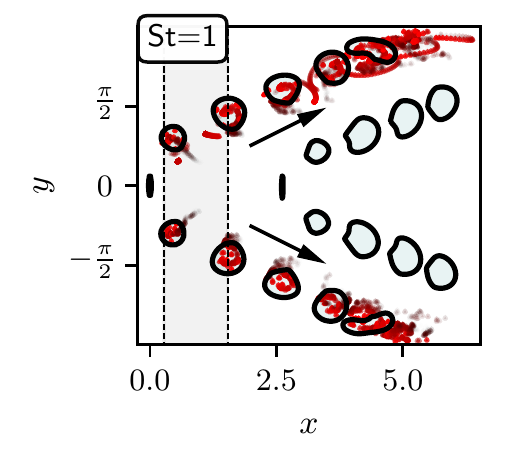}
        \includegraphics[width=0.28\textwidth,trim=32.25 0 0 0, clip]{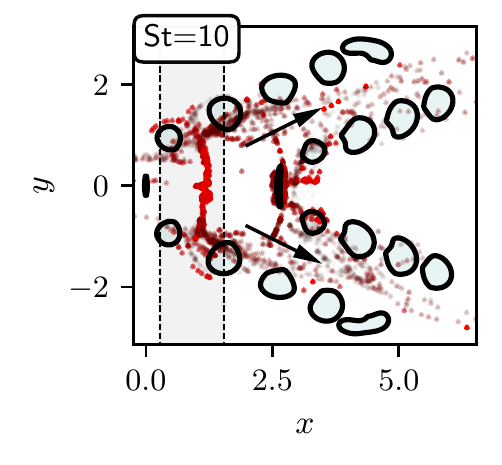}
    \caption{Top row: Time evolution of mean streamwise particle position (left) and mean wall-normal particle position within the channel (right) with varying $St$. The shaded bands indicate one standard deviation. Bottom row: Particle location at six time instances (separated by a $dt=10$) for different $St$ numbers. The contours correspond to the Q-criteria (-0.02) of the centerplane. Particle location is overlaid and the position in the $z$-direction is shown with transparency of the symbols. The reconnection events occur within the shaded region}
    \label{fig:particlePosition}
    \label{fig:meanParticlePosition}
\end{figure}

 The relative particle to fluid timescale, characterised by the particle Stokes number,  determines the exact trajectory of the transported particles.  The mean particle position during the transport and reconnection is shown in figure~\ref{fig:meanParticlePosition} (top, left). For sufficiently small Stokes number ($St=$ 0.1 and 1), the particles are maintained within the orbit of the vortex core and undergo a long streamwise transport due to the self-advection of the rings; in the limit of $St\rightarrow 0$ would identically follow the pathlines of the flow.  As the rings undergo two sequential reconnection events---ultimately resulting in two new vortices---the low-Stokes-number particles stay within the newly-formed vortex rings and are transported towards the sidewalls. The particle trajectories of cases  $St=0.1$ and 1 differ in that the heavier particles $St=1$ get transported away from central vortex line and rotate around it (as evidenced by the oscillatory mean streamwise position of $St=1$ between $t=5$ and 20). Despite this, the particles remain within the vortex core as shown in figure \ref{fig:meanParticlePosition} (bottom, middle). The particle motion within the vortex core does not significantly affect the deposition location.
 For higher Stokes ($St=10$), the heavier particles are unable to be transported by the vortex ring and get dispersed in the channel as noted by the large standard deviation of the particle location.



\begin{figure}
\centering
   \includegraphics[width=0.8\textwidth]{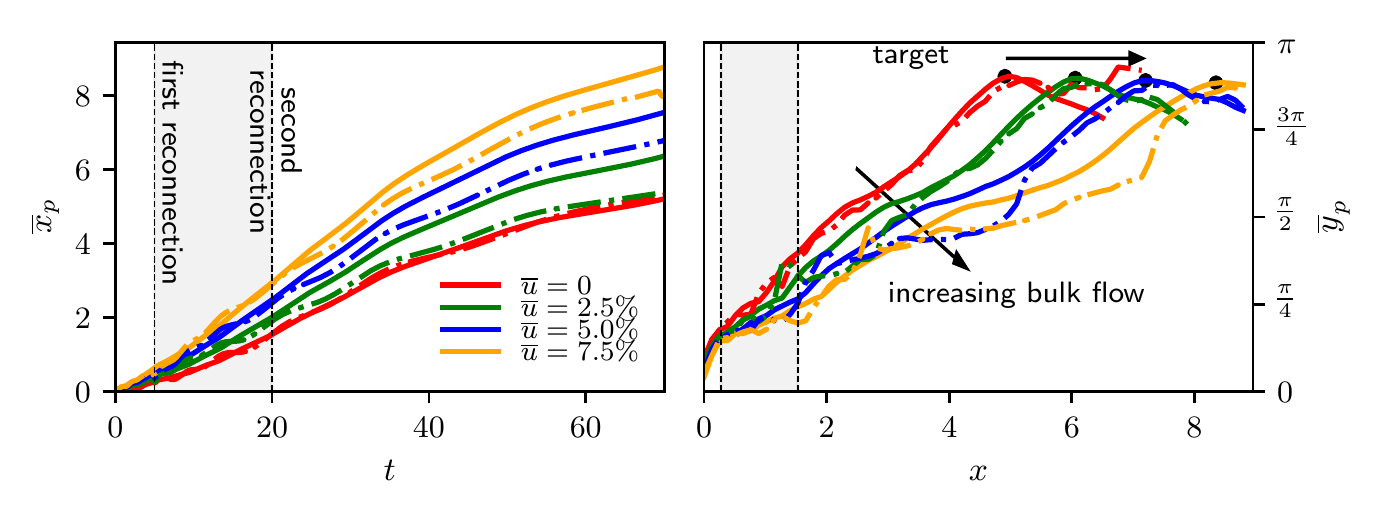}\vspace{-0.95cm}
       \includegraphics[width=0.8\textwidth]{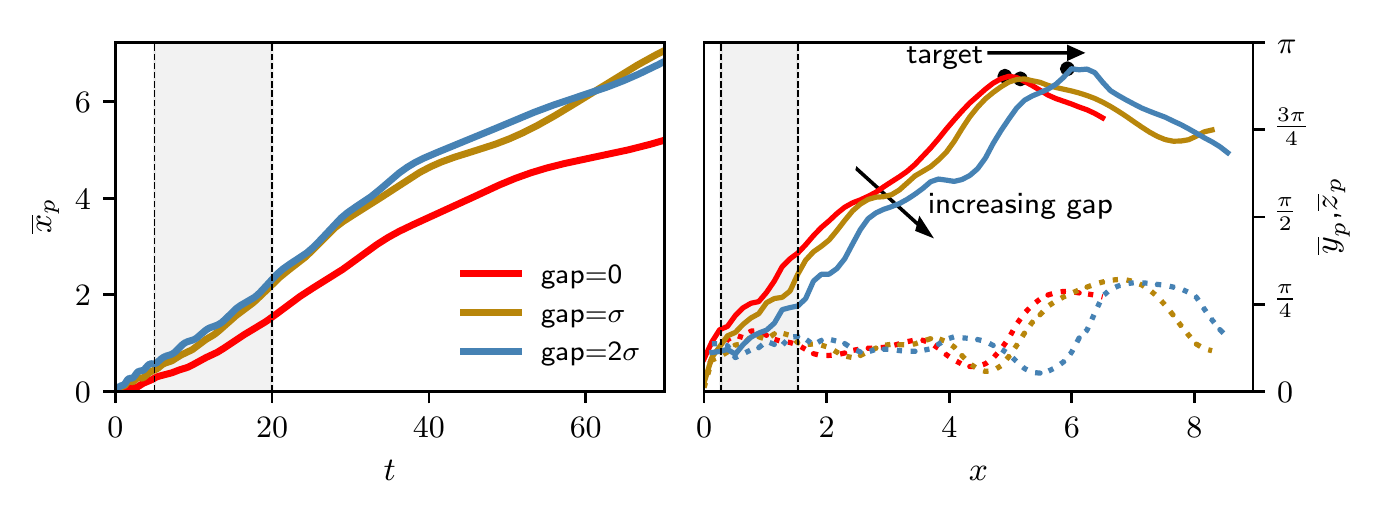}
    \caption{Evolution of  streamwise particle position (left) and  wall-normal particle position  (right). Top row: varying bulk channel velocity at $St=0.1$ (dash-dot line) and $St=1$ (full line). Bottom row:  varying initial gaps; $\sigma$ is the vortex core radius (right, full lines: $\overline{y}_p$, dotted lines: $\overline{z}_p$). Shaded regions indicate the approximate location of vortex reconnection}
    \label{fig:vortex_bulkFlow}
\end{figure}

The control of the particle deposition location is now considered.  Due to the low-Reynolds number of the simulations and the modest circulation of the vortex rings, the particles have insufficient wall normal momentum to impact the sidewalls. As a result, we consider the particles  to be ``delivered'' at the location at which the average distance to the wall is minimized.  The parameter controlling the location of the particle deposition is tied to the location and the  strength of the reconnection event. The location of reconnection is affected by the background flow and the separation distance of the side-by-side vortex rings. On the other hand, the strength of reconnection, which is tied to the circulation-based Reynolds number,  velocity of the particles in the streamwise (prior to reconnection) and wall-normal  (post reconnection)  direction. The addition of a laminar,  bulk flow within the channel does not disrupt the reconnection events as long as the bulk velocity is low. It is observed that a bulk channel flow velocity above about 7.5\% of the maximum velocity of the stationary vortex rings impacts the reconnection.  As  expected, there is an increased advective particle velocity with increasing bulk channel flow, this results in a farther deposition location within the channel, as shown in  figure \ref{fig:vortex_bulkFlow} (top). As the particle motion is now due to the combined effects of the vortex self-advection and bulk flow transport, an increase of the bulk channel flow velocity modifies the effective particle Stokes number. The results in figure \ref{fig:vortex_bulkFlow} (top, right) shows a linear increase in the location of particle deposition; from $x=4.9$ without a base flow, to $x=8.35$ with a 7.5\% bulk velocity.
An increase in the initial separation distance between the co-planar vortex rings, defined by $\sigma$, causes  a longer vortex self-advection period prior to reconnection. As shown in figure \ref{fig:vortex_bulkFlow} (bottom, left), this results in a drastically faster particle advection in the channel and the delayed reconnection causes the deposition site to occur farther downstream, figure \ref{fig:vortex_bulkFlow} (bottom, right).



\section{Conclusion}
We propose a conceptual approach to target particle delivery towards the sidewall of internal flows via particle-laden vortex ring reconnection and show the feasibility of the approach using two-way coupled, Eulerian-Lagrangian point-particle simulations. Particle-loaded  vortex rings undergo a self-induced translational motion. If a pair of co-planar vortex rings touch, they will undergo a viscous reconnection process which leads to the pinching off of two newly vortex rings. Throughout reconnection, the initially streamwise translating rings particles  get re-directed towards the sidewall of internal flow. The ability to deliver particles trapped in the core of the vortex ring is affected by the particle Stokes number and can be tuned by modifying either the location or strength of the reconnection event. In this work, we focus on the modification of the reconnection location by showing the effects of initial rings' separation and bulk flow velocity. In future works, low-order analytical models will be established for targeted particle delivery.

\bibliographystyle{apalike}
\bibliography{main.bib}

\end{document}